\documentclass{acm_proc_article-sp}

\begin{document}
\title{A Study and Implementation of  {\ttlit RSA} Cryptosystem
\titlenote{Permission to make digital or hard copies of all or part of this work for personal or classroom use is granted without fee provided that copies are not made or distributed for profit or commercial advantage and that copies bear this notice and the full citation on the first page.}}
\subtitle{[Implementing the well known Rivest Shamir Adleman Public Key Cryptosystem]
\titlenote{A full code of this implementation is availble at
\texttt{{https://github.com/lugju/rsa}}}}
%
%
%
%
%

\numberofauthors{2} 
%
\author{
%
%
\alignauthor
Sinjan Chakraborty\\
       \affaddr{Computer Science and Engineering Department}\\
       \affaddr{Jadavpur University}\\
       \affaddr{Kolkata, 700032 | WB IN}\\
       \email{sinjanc@gmail.com}
\alignauthor
Vineet Kumar\titlenote{Both author of this paper are freshman students auditing the \textit{RC Bose Summer Research Internship Program} on Cryptography at \textit{Indian Statistical Institute, Calcutta}. Both authors are members of {Linux User's Group Jadavpur University}. }\\
       \affaddr{Computer Science and Engineering Department}\\
       \affaddr{Jadavpur University}\\
       \affaddr{Kolkata, 700032 | WB IN}\\
       \email{vntkumar8@gmail.com}
}

\maketitle
\begin{abstract}
This project involves an implementation of the Rivest Shamir Adleman (RSA)\cite{rsa} encryption algorithm in C. It consists of generation of two random prime numbers and a number co-prime to \begin{math} \phi(n) \end{math} also called euler toitent function. These three are used to generate a public key and private key. The user has to enter a message which is encrypted by the public key. The algorithm also decrypts the generated cipher text with the help of the private key and returns the plain text message which was encrypted earlier.
\end{abstract}


\terms{}
Plain Text: The general message needed to be encrypted is called plain text.\\
Cipher Text: The Garbage like looking string having no information about plain text is called cipher text.
\keywords{RSA, Public Key Cryptosystem, Asymmetric Key Cryptosystem, Cryptography, Security.} 

\section{Introduction}
This project has been done so that people interested in cryptography, especially students can get a better understanding of how generation of random primes, key generation and encryption and decryption process takes place in RSA. Being an undergrad students our search for an easy-to-understand C implementation of RSA that involved generation of random primes lead no where. On not finding a satisfactory implementation, we implemented it.    
\section{Public-Key Cryptosystems}
In a \textit{public key cryptosystem} each user places in a public file an encryption procedure E. That is, the public file is a directory giving the encryption procedure of each user. The user keeps secret the details of his corresponding decryption procedure D. These procedures have the following four properties:
\begin{enumerate}
	\item Deciphering the enciphered form of a message M yields M. Formally,
	
		\[D(E(M) = M
	\]
	\item Both E and D are easy to compute
	\item By publicly revealing E the user does not reveal an easy way to compute D. This means that in practice only he can decrypt messages encrypted with E, or compute D efficiently
	\item If a message M is first deciphered and then enciphered, M is the result. Formally,
	
		\[E(D(M) = M
	\]
	
\end{enumerate}

\section{Implementation {\secit RSA} Algorithm}
For Implementation of RSA these four famous algorithms have been used: -
\begin{itemize}
	\item Sieve of Eratosthenes (for prime number generation)
	\item Fermat Primality Test
	\item Miller-Rabin Primality Test 
	\item Encryption-Decryption Algorithms.
\end{itemize}
 Proper discussion about each algorithm is mentioned in next subsections.

\subsection{Sieve of Eratosthenes}
The sieve of Eratosthenes is one of the most efficient ways to find all primes smaller than n when n is smaller than 10 million or so. Following is the algorithm to find all the prime numbers less than or equal to a given integer n by Eratosthenes method:

1.	Create a list of consecutive integers from 2 to n : (2, 3, 4,.... , n).

2.	Initially, let p equal 2, the first prime number.

3.	Starting from p, count up in increments of p and mark each of these numbers greater than p itself in the list. These numbers will be 2p, 3p, 4p, etc.; note that some of them may have already been marked.

4.	Find the first number greater than p in the list that is not marked. If there was no such number, stop. Otherwise, let p now equal this number (which is the next prime), and repeat from step 3.
But When the algorithm terminates, all the numbers in the list that are not marked are prime.

\subsection{Fermat's Primality Test}
If p is the number which we want to test for primality, then we could randomly choose a, such that \textit{a < p} and then calculate \begin{math}
	a^{(p-1)} modulo
\end{math}\textit{ p}. If the result is not 1, then by Fermat's Little Theorem \textit{p} cannot be prime. What if that is not the case? We can choose another \textit{a} and then do the same test again. We could stop after some number of iterations and if the result is always 1 in each of them, then we can state with very high probability that \textit{p} is prime. The more iterations we do, the higher is the probability that our result is correct. You can notice that if the method returns composite, then the number is sure to be composite, otherwise it will be probably prime.

\begin{verbatim}
int Fermat(long long p,int iterations){
    if(p == 1){ /*1 isn't prime; this conditional 
		statement is not required for our given program. 
		We write it for maintaining generality.*/
        return 0;                       
    }   
    int i=0;            
    for(i=0;i<iterations;i++){
        /* choose a random integer between 
				1 and p-1 ( inclusive )*/
        long long a = rand()%(p-1)+1; 
        /* modulo is the function we developed 
				above for modular exponentiation */
        if(modulo(a,p-1,p) != 1){ 
            return 0; /* p is definitely composite */
        }
    }
    return 1; /* p is probably prime */
}

\end{verbatim}
\subsection{Miller-Rabin Primality Test }
Let \textit{p} be the given number which we have to test for primality. First we rewrite \textit{p-1} as \begin{math}
	2d*s
\end{math} . Now we pick some \textit{a} in range [\textit{1,n-1}] and then check whether \textit{a*s = 1 ( mod p )} or \begin{math}
	a*s*2r = -1 ( mod p )
\end{math}. If both of them fail, then \textit{p} is definitely composite. Otherwise \textit{p} is probably prime. We can choose another \textit{a} and repeat the same test. We can stop after some fixed number of iterations and claim that either \textit{p} is definitely composite, or it is probably prime. It can be shown that for any composite number \textit{p}, at least (3/4) of the numbers less than \textit{p} will witness \textit{p} to be composite when chosen as \textit{a} in the above test. Which means that if we do 1 iteration, probability that a composite number is returned as prime is (1/4). With k iterations the probability of test failing is \textit{(1/4)k} or \textit{4(-k)}. This test is comparatively slower compared to Fermat's test but it doesn't break down for any specific composite numbers and 18-20 iterations is a quite good choice for most applications.

\begin{verbatim}
int Miller(long long p,int iteration){
    if(p<2){ /*this conditional statement is 
		not required for our given program. 
		We write it for maintaining generality.*/
        return 0;
    }
    if(p!=2 && p%2==0){
        return 0;
    }
    long long s=p-1;
    while(s%2==0){
        s/=2;
    }
    int i=0;
    for(i=0;i<iteration;i++){
        long long a=rand()%(p-1)+1,temp=s;
        long long mod=modulo(a,temp,p);
        while(temp!=p-1 && mod!=1 && mod!=p-1){
            mod=mulmod(mod,mod,p);
            temp *= 2;
        }
        if(mod!=p-1 && temp%2==0){
            return 0;
        }
    }
    return 1;
}
\end{verbatim}
\subsection{RSA Algorithm}
The algorithm of RSA consists of the following steps:

1.	Select two prime nos \textit{ p \& q} such as \textit{p!=q}

2.	Calculate \textit{n} as product of \textit{p \& q}, i.e. \textit{n=p*q}

3.	Calculate \textit{m} as product of   \textit{(p-1)   \&   (q-1)} i.e.          \textit{m=(p-1)(q-1)}

4.	Select any integer \textit{e<m} such that it is co-prime to \textit{m},  co-prime  means  gcd(e,m) = 1.

5.	Calculate \textit{d} such that \textit{(d*e) mod m = 1} ,                 \\ i.e. \textit{d = e-1  mod m}

6.	The public key is {\textit{e,n}} The private key is {\textit{d,n}}
\\So these are the keys, now if you want to perform some encryption operation using these keys here are the steps, \\
if you have a text \textit{P}, its encrypted version(cipher text \textit{C} is)
\begin{math}
	C= P^e  mod n
\end{math}                                                                               \\ To decrpty it back to plain text use
\begin{math}
	P= C^d  mod n
\end{math} 

\section{Implementation}
Successful Implementation of above discussed algorithms can be achieved in following steps. 
\subsection{Generating Two Large Primes}
In the implementation of RSA, the first thing we have to do is generate two large prime numbers, \textit{p} and \textit{q}. The standard way to generate big prime numbers is to take a preselected random number of the desired length, apply a Fermat Test (best with the base 2 as it can be optimized for speed) and then to apply a certain number of Miller-Rabin tests (depending on the length and the allowed error rate like \begin{math}
	2^{-100}
\end{math} ) to get a number which is very probably a prime number. The pre-selection is done either by test divisions by small prime numbers (up to few hundred) or by sieving out primes up to 10,000 - 1,000,000 considering many prime candidates of the form \textit{b+2i}(\textit{b} big, \textit{i} upto a few thousands). 

Here pre selection of a random prime by first creating an array of primes of size 10000 by the Sieve of Eratosthenes\footnote{all numbers in this array created by sieving may not be prime}. Then generated a random number `gen' with the help of \texttt{rand()} function  such that the number is greater than 1000 and less than 10000. I have extracted the number at `gen'-th index position of the array of primes\footnote{It is customary to use numbers of this range because implementation is done on a \texttt{INTEL(R) CORE\begin{math}
	^{TM}
\end{math}i5-4210U CPU @ 1.7 GHz 2.4 Ghz}. Otherwise if not confined in given range, it was taking a lot of time to decrypt a cipher. For this range, the probability that a number in the array is not a prime is really very low. But as we increase the upper bound of these numbers, this probability increases. So applying Fermat Primality Test and Miller-Rabin Primality Test on the number is very essential.}.

Now as the extracted number might not be a prime number, applied the Fermat Primality Test. Though Fermat is highly accurate in practice there are certain composite numbers \textit{p} known as \textit{Carmichael numbers} for which all values of \textit{a<p} for which \textit{gcd(a,p)=1} .\begin{math}
	(a(p-1))
\end{math}\textit{modulo p =1}. If  apply Fermat's test on a Carmichael number the probability of choosing an a such that \textit{gcd(a,p) != 1} is very low ( based on the nature of Carmichael numbers), and in that case, the Fermat's test will return a wrong result with very high probability.

For this reason, applied Miller Rabin Primality Test on the number in case it satisfies Fermat test. If the number satisfies all the tests, we get a large number with a very high probability of being prime. 

Now  put the entire set of above operations in a loop such that if the number generated does not satisfy the Miller Rabin Primality Test, the above operations will be repeated again and this will go on until we generate a prime number. As needed to generate two prime numbers, we provide another loop outside this which runs two times and stores the generated primes in an array of size 2.
\subsection{Key Generation}
Calculate \textit{n} as the product of \textit{p \& q}, the two generated primes. The next thing to do is to generate a number \textit{e} such that ${e<\phi(n)}$ and $gcd(e,\phi(n))=1$. Here generated a random number less than $\phi(n)$ and checked their gcd. Continued the process until their gcd is 1. 
After generating \textit{e},Calculated the value of \textit{d} such that \begin{math}
	d = e^{-1 } mod \phi(n).
\end{math}
The public key is \textit{\{e,n\}}.
The private key is \textit{\{d,n\}}.
\subsection{Encryption}
Created a function \texttt{encrypt()} to encrypt a message with the help of public key \textit{\{e,n\}}. The encrypted  keyword \textit{C} is calculated by doing \begin{math}
	C=M^e  mod n
\end{math}.

\begin{verbatim}
void encrypt(){
      long long i;
      C = 1;
      for(i=0;i< e;i++)
      C=C*M%n;
      C = C%n;
      printf("\n\tEncrypted keyword : %lld",C);
}
\end{verbatim}
\subsection{Decryption}
Created a function \texttt{decrypt()} to decrypt a cipher with the help of private key\textit{\{d,n\}}.  The decrypted message \textit{M} is calculated by doing 
\begin{math}
	M= C^d  mod n
\end{math}.
\begin{verbatim}
void decrypt(){
      long long i;
      M = 1;
      for(i=0;i< d;i++)
            M=M*C%n;
            M = M%n;
      printf("\n\tDecrypted keyword : %lld\n",M);
}
\end{verbatim}
\begin{table}

\caption{Data Types Used in Implementation}
\begin{tabular}{|c|c|l|} \hline
Variable Name&DataType&Use\\ \hline
     p,q &	long long	& Stores the two generated primes\\
 n &	long long	&Stores product of p and q.\\
 M	&long long&	Stores the message entered \\
 phi&	long long&	Stores value of  $\phi (n) $=(p-1)(q-1)\\
 e	&long long&	Stores the public exponent e\\
 d&	long long&	Stores the private exponent d\\
     C&	long long&	Stores the Cipher\\

\hline\end{tabular}
\end{table}
	\section{Simple Examples}
	Example 1:
	\begin{verbatim}
62011  
12269 
 
	 F(n) phi value 	= 760738680 
	Public Key	: {11723299,760812959} 
	Private Key	: {288096259,760812959} 
 
Enter The Plain Text	: 5321 
 
	Encrypted keyword : 573183424 
 
Enter the Cipher text	: 573183424 
 
	Decrypted keyword : 5321 

	\end{verbatim}
	
	Example 2:

	\begin{verbatim}
	39703  
66883  
 
	 F(n) phi value 	= 2655349164 
	Public Key	: {8068769,2655455749} 
	Private Key	: {149069429,2655455749} 
 
Enter The Plain Text	: 5321 
Encrypted keyword : 2521426694 
 
Enter the Cipher text	: 2521426694 
 
	Decrypted keyword : 5321
	\end{verbatim}
	
	\textsc{Comment:} These two examples show how on entering the same message on two different runs, the value of the cipher text is different because of the generation of different prime numbers \textit{p} and \textit{q} and the different public and private exponents \textit{e} and \textit{d}, each time the program is run.

\section{Problems in Implementation}
The main problem faced is that for a given message \textit{M}, the cipher text \textit{C} generated is large, nearly a 10 digit number. Also, the value of \textit{d}, i.e., the private exponent is of 10 digits. To decrypt \textit{M}, we have to calculate $M=C^d  mod n$. This calculation is taking a long time in some cases. 
\section{Analysis}
The time taken on computing is not fixed it changes in each \& every run. After performing a statistical test of encrypting a plain text (which is 25000) \& decrypting it time taken is noted 

\begin{table}[h]
	\centering
		\begin{tabular}
					{|l|l|}\hline
		\textbf{Value of F(n)}&	\textbf{Time (in sec.)}\\ \hline
1614044880&	16.5\\ \hline
946335840&	21.5\\ \hline
2186888896&	60.0\\ \hline
1927086672&	17.6\\ \hline
1595209116&	18.8\\ \hline
1424163096&	9.4\\ \hline
2917216608&	28.2\\ \hline

		\end{tabular}
	\caption{Computational Complexity (Data)}
	\label{tab:ComputationalComplexityData}
\end{table}
Plotting the data it is extremely clear that computational time of phi function, encryption \& decryption of plain text \& cipher text respectively is not predictable. Some runs of test with takes more time,  some takes less time while complexity of computations remaining approximately fixed. 

\begin{figure}[h]
	\centering
		\includegraphics[width=0.50\textwidth]{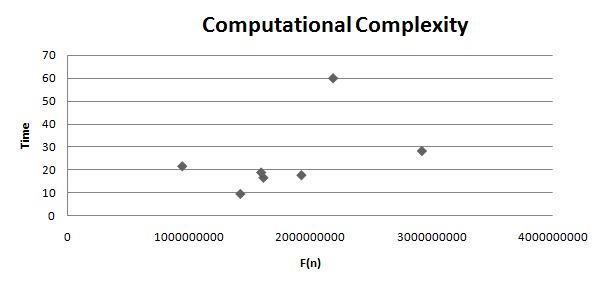}
	\caption{Computational Complexity (Graph)}
\end{figure}

%

\balancecolumns
\end{document}